\documentclass[conference,a4paper]{IEEEtran}
\usepackage[left=1.43cm,right=1.43cm,top=1.8cm,bottom=4.21cm]{geometry}
\IEEEoverridecommandlockouts
\usepackage{cite}
\usepackage{amsmath,amssymb,amsfonts,bbm}
\usepackage{graphicx}
\usepackage{textcomp}
\usepackage[dvipsnames]{xcolor}
\usepackage{multirow}
\usepackage{subcaption}
\usepackage{caption}\usepackage{tikz}
\usepackage{tkz-tab}
\usetikzlibrary{automata,arrows,positioning,calc}
\usetikzlibrary{shapes,snakes}
\usetikzlibrary{arrows}
\usepackage{dsfont}
\usepackage{soul}
\usepackage{subfiles}
\usepackage{comment}
\usepackage{algorithm}
\usepackage{algpseudocode}
\usepackage{amsmath}
\usepackage{amssymb}
\usepackage{amsthm}
\usepackage{booktabs}
\usepackage{epstopdf} 

\def\BibTeX{{\rm B\kern-.05em{\sc i\kern-.025em b}\kern-.08em
    T\kern-.1667em\lower.7ex\hbox{E}\kern-.125emX}}

\newcommand{\hlc}[2][yellow]{{%
    \colorlet{foo}{#1}%
    \sethlcolor{foo}\hl{#2}}%
}

\usepackage{tikz}
\usetikzlibrary{fit,calc}
\newcommand*{\tikzmk}[1]{\tikz[remember picture,overlay,] \node (#1) {};\ignorespaces}
\newcommand{\boxit}[1]{\tikz[remember picture,overlay]{\node[yshift=-10pt,xshift=-130pt, fill=#1,opacity=.25,fit={(A)($(B)+(1.12\linewidth,.8\baselineskip)$)}] {};}\ignorespaces}
\colorlet{pink}{red!40}
\colorlet{blue}{cyan!50}

\begin{document}

\bstctlcite{IEEEexample:BSTcontrol}

\title{ConPA: A Contention-free Mechanism with\\Power Adaptation for Beyond Listen-Before-Talk\\
\thanks{The work of P. Baracca was supported by the German Federal Ministry of Education and Research (BMBF) project 6G-ANNA grant 16KIS077K.}
}

\author{
\IEEEauthorblockN{Francesc Wilhelmi$^{\star}$, Paolo Baracca$^{\flat}$, Gianluca Fontanesi$^{\star}$, and Lorenzo Galati-Giordano$^{\star}$ \vspace{0.1cm}
}
\IEEEauthorblockA{$^{\star}$\emph{Radio Systems Research, Nokia Bell Labs, Stuttgart, Germany}}
\IEEEauthorblockA{$^{\flat}$\emph{Nokia Standards, Munich, Germany}}
}

\maketitle

\thispagestyle{plain}
\pagestyle{plain}

\begin{abstract}
In view of the need to find novel means to utilize the unlicensed spectrum to meet the rising latency and reliability requirements of new applications, we propose a novel mechanism that allows devices to transmit anytime that a packet has to be delivered. The proposed mechanism, Contention-free with Power Adaptation (ConPA), aims to bypass the contention periods of current Listen-Before-Talk (LBT) approaches, which are the main source of unreliability in unlicensed technologies like Wi-Fi. To assess the feasibility of ConPA, we provide an analytical method based on Markov chains, which allows deriving relevant performance metrics, including throughput, airtime, and quality of transmissions. Using such a model, we study the performance of ConPA in various scenarios, and compare it to baseline channel access approaches like the Distributed Coordination Function (DCF) and the IEEE 802.11ax Overlapping Basic Service Set (OBSS) Packet Detect (PD)-based Spatial Reuse (SR). Our results prove the effectiveness of ConPA in reusing the space to offer substantial throughput gains with respect to the baselines (up to 76\% improvement).
\end{abstract}


\section{Introduction}
\label{sec:introduction}

Future wireless communications systems will have to support the extreme performance requirements of next-generation applications like industrial digital twinning or immersive telepresence~\cite{uusitalo20216g}, which emphasize latency and reliability atop of throughput, capacity, and coverage~\cite{berardinelli2021extreme}. A significant portion of these systems is expected to operate in the unlicensed spectrum, which is regarded as an appealing option for easy deployment of private solutions and free access to a vast chunk of frequency bands (e.g., up to 1.2~GHz of spectrum in the 6~GHz band). However, most of the operations in unlicensed bands below 7 GHz are regulated by Listen-Before-Talk (LBT), a mechanism based on Carrier Sense Multiple Access (CSMA) with Collision Avoidance (CA) that was purposed to provide fair distributed channel access. The LBT procedure relies on Clear Channel Assessment (CCA), whereby a transmitter device must first perform Energy Detection (ED) to check if the measured power at the intended transmission channel is above a certain threshold (i.e., the channel is busy thus deferring transmission) or not (i.e., the channel is idle thus allowing transmission). 

Such a spontaneous and distributed manner of using the unlicensed spectrum has, however, obvious limitations. LBT guarantees co-existence and fairness by sacrificing latency as a requirement, which makes LBT-based channel access techniques not particularly suitable for applications with stringent requirements. While communication protocols operating in the unlicensed spectrum like IEEE 802.11 (aka Wi-Fi) are evolving towards meeting reliability and low-latency targets (e.g., through new functionalities enforcing coordination~\cite{wilhelmi2023throughput, GalGerCar2023}), novel channel access rules are attracting attention and triggering research debates for their adoption in future available unlicensed spectrum~\cite{fcc2023fact}. 

Several proposals have tried to enhance channel access in the unlicensed by tweaking LBT in multiple ways (e.g., applying spatial filtering) and for multiple purposes (e.g., high reliability, low latency, spectral efficiency). In this regard, interference management has been shown to be critical to overcoming the limitations of LBT~\cite{akella2005self}. The most prominent state-of-the-art mechanism in this area is Overlapping Basic Service Set (OBSS) Packet Detect (PD)-based Spatial Reuse (SR), adopted in the 802.11ax amendment~\cite{wilhelmi2021spatial} to provide further efficiency in sharing the unlicensed medium through power control and interference sensitivity adjustment. Other solutions like~\cite{ma2008joint, afaqui2016dynamic} also proposed to tune the transmit power and/or the Physical Carrier Sense (PCS) threshold to enhance spatial reuse. Alternatively, a dynamic PCS threshold method aided by Artificial Intelligence (AI) was proposed in~\cite{ak2020fsc}. 

The prior art has attempted to improve the current LBT operation but without substantially going beyond its contention principles. In this paper, we propose a new distributed channel access method for the unlicensed spectrum, named Contention-free mechanism with Power Adaptation (ConPA). Unlike current existing methods based on LBT, ConPA allows devices to access the channel when required, thus aiming at providing increased performance and higher reliability. To do so, ConPA removes channel contention, and its associated unreliability, and adapts the transmission power to be used based on the interference measured during ED, such that higher transmit power is allowed when lower interference is measured. The main contributions of this paper are as follows:
\begin{itemize}
    \item We propose ConPA, an airtime-friendly mechanism based on power adaptation for distributed channel access.
    \item We provide an analytical model based on Continuous Time Markov Chains (CTMCs) to analyze ConPA.
    \item We evaluate the proposed ConPA mechanism and compare its performance against baseline LBT-based methods, including Distributed Coordination Function (DCF) and IEEE 802.11ax OBSS/PD-based SR.
\end{itemize}

\section{ConPA Design}
\label{sec:channel_access}

\begin{figure}[ht!]
    \centering    
    \includegraphics[width=\columnwidth]{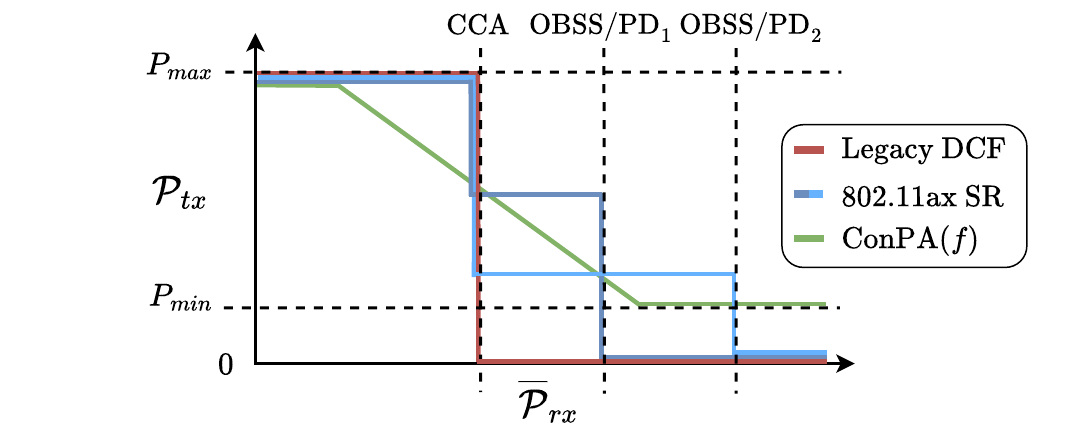}
    \caption{Power mapping functions.}
    \label{fig:power_mapping}
\end{figure}

We start describing the channel access rules of the baseline approaches considered in this paper, namely the IEEE 802.11 DCF (referred to as \textit{DCF}) and IEEE 802.11ax OBSS/PD-based SR (referred to as \textit{802.11ax SR}). Then, we describe the proposed ConPA mechanism, followed by its characterization through CTMCs. As a way of summary, Fig.~\ref{fig:power_mapping} illustrates the power mapping policies applied by each mechanism, where the allowed power for transmission is represented as a function of the measured interference. The allowed power may become zero in the baseline approaches, thus resulting in deferring the transmission. ConPA, in contrast, ensures consistent channel access with a minimum transmit power.

\subsection{DCF and 802.11ax SR}

In Wi-Fi, channel access is based on the DCF, being the exponential backoff procedure the main primitive for addressing decentralization. In particular, a Wi-Fi device with packets in the buffer will initiate a transmission if the channel has been detected idle during a backoff period, computed randomly based on a Contention Window (CW) parameter. In the DCF channel access, the channel is marked as idle as long as no detected Wi-Fi signal is above a given CCA threshold (CCA$_{thr}$). In 802.11ax SR, in contrast, a more aggressive channel access policy is implemented to reduce contention. In that case, provided that the channel has been detected busy due to an inter-BSS transmission, an additional threshold is applied, namely the OBSS/PD threshold (OBSS/PD$_{thr}$), with OBSS/PD$_{thr}$ $\geq$ CCA$_{thr}$. The OBSS/PD$_{thr}$ comes along with a transmit power limitation and the duration of SR-based TXOPs is limited by the duration of the firstly detected transmissions. Both DCF and 802.11ax SR channel access mechanisms are briefly described in Alg.~\ref{alg:dcf_11ax}. The specific parts of 802.11ax SR are highlighted in \hlc[NavyBlue!30]{blue}, while the rest of the steps are common for both DCF and 802.11ax SR.

\algdef{SE}[SUBALG]{Indent}{EndIndent}{}{\algorithmicend\ }%
\algtext*{Indent}
\algtext*{EndIndent}
\begin{algorithm}[t!]
\caption{\hlc[pink!30]{DCF} and \hlc[NavyBlue!30]{802.11ax SR} channel access}\label{alg:dcf_11ax}
\begin{algorithmic}[1]
    \State \textbf{Initialize:} CW, $\text{CCA}_{thr}$, \hlc[NavyBlue!30]{$\text{OBSS/PD}_{thr}$}, \hlc[NavyBlue!30]{$\mathcal{P}_{rx}^{ref}$}
    \While {Backoff ($BO$) is active}
    \State \textbf{on} \texttt{Inter-BSS-signal-detect}($\mathcal{P}_{rx}$):
    \Indent
       \If{$\mathcal{P}_{rx} < \text{CCA}_{thr}$} 
            \State \text{CHANNEL\_STATUS} $\leftarrow$ \text{IDLE}
            \State $BO \leftarrow BO - 1$
        \Else
           \State \text{CHANNEL\_STATUS} $\leftarrow$ \text{BUSY}
            \tikzmk{A}
            \If{$\mathcal{P}_{rx} < \text{OBSS/PD}_{thr}$}
                \State \text{CHANNEL\_STATUS} $\leftarrow$ \text{IDLE\_SR}
                \State $BO \leftarrow BO - 1$
            \EndIf
            \tikzmk{B}
            \boxit{NavyBlue!130}
        \EndIf
    \EndIndent
    \EndWhile
    \If{Backoff ($BO$) expires}
        \State \textbf{on} \texttt{packet-transmission}(TXOP-type):
        \Indent
           \If{TXOP-type == SR\_TXOP} 
           \State \textproc{\hlc[NavyBlue!30]{Adjust Transmit Power}}\hlc[NavyBlue!30]{(OBSS/PD$_{thr}$)}
           \EndIf
        \EndIndent
    \EndIf
    \Procedure{Adjust Transmit Power}{}
        \State \hlc[NavyBlue!30]{$\mathcal{P}_{tx} = \mathcal{P}_{rx}^{ref} - (\text{OBSS/PD}_{thr} - \text{OBSS/PD}_{min})$} 
    \EndProcedure         
\end{algorithmic}
\end{algorithm}

\subsection{ConPA}

\begin{algorithm}[t!]
\caption{ConPA using \hlc[ForestGreen!40]{$f$}}\label{alg:conpa} 
\begin{algorithmic}[1]
    \State \textbf{Initialize:} CW,  $\mathcal{P}_{min}$, C
    \While {Backoff ($BO$) is active}
    \State \textbf{on} \texttt{Inter-BSS-signal-detect}($\mathcal{P}_{rx}$):            
        \State \indent $\mathcal{P}_{rx}^{HIST} \leftarrow \mathcal{P}_{rx}^{HIST}\cup \mathcal{P}_{rx}$
        \State \indent $BO \leftarrow BO - 1$
    \EndWhile
    \If{Backoff ($BO$) expires}
    \State \textbf{on} \texttt{packet-transmission}($\mathcal{P}_{rx}^{HIST}$):
         \State \indent $\mathcal{\overline{P}}_{rx} \leftarrow \frac{1}{N} \sum_{i=0}^{N-1}\mathcal{P}_{rx}^{HIST}(n-i)$
        \State \indent \Call{Adjust Transmit Power}{$\mathcal{\overline{P}}_{rx}$, $f$}  
    \EndIf
    \Procedure{Adjust Transmit Power}{}
        \State \hlc[ForestGreen!40]{$f(\mathcal{\overline{P}}_{rx})$}$: \mathcal{P}_{tx} = \max\{\mathcal{P}_{min}, \text{C} - \mathcal{\overline{P}}_{rx}\}$
    \EndProcedure         
\end{algorithmic}
\end{algorithm}

Using ConPA, since contention periods are bypassed, a device can always transmit provided that it adjusts the transmit power as a function of the sensed interference. While limiting SINR due to potentially increased interference regimes, our approach targets increased airtime, which can contribute to both achieve stringent latency requirements and provide higher throughput. Algorithm~\ref{alg:conpa} describes the basic operation of ConPA with an interference-to-power mapping function \hlc[ForestGreen!40]{$f$} that uses the power of the sensed interference and a constant $\text{C}$ (notice that other transmit power adaptation functions could be used instead). ConPA takes power measurements during the backoff time so that any new signal detected contributes to filling a registry $\mathcal{P}_{rx}^{HIST}$ (line 4). When the backoff countdown expires, the transmission is subject to the applied transmit power adjustment function (line 13), which uses the last $N$ measurements in $\mathcal{P}_{rx}^{HIST}$ to estimate $\overline{P}_{rx}$ (lines 7-11). In this work, we assume that $\overline{P}_{rx}$ is statically defined by the aggregate power from receiver devices involved in active transmissions.

\subsection{CTMC Characterization}

CTMCs characterize feasible states (representing feasible combinations of ongoing transmissions, determined by the channel access mode), forward transitions (ruled by the channel access attempt rate, $\lambda$), and backward transitions (ruled by the channel departure rate, $\mu$). The steady-state probability of each state in the chain, characterized by $\pi_s$, is obtained by solving $\pi_s Q = 0$, where $Q$ is the transitions rate matrix, filled with both forward and backward transitions. More specifically,
\begin{itemize}
    \item Forward transitions are defined by the channel access attempt rate, which is based on the expected backoff time $E[BO] = 2/(\text{CW}-1) \, T_{\textrm{e}}$ (where $T_{\textrm{e}}$ is the duration of an empty slot). The channel access rate is defined as $\lambda = \alpha/E[BO]$. To characterize non-saturated traffic, we rely on the long-term probability of having packets ready when the backoff expires, as done in~\cite{barrachina2019overlap}. As a result, the channel access rate is multiplied by a scalar $\alpha \in [0,1]$.
    \item Backward transitions are ruled by the time a BSS is active once it gains channel access, which depends on the time for transmitting the data and the associated control packets, $T$. Backward transitions are thus characterized by a parameter $\mu$, which is obtained as $1/T$.
\end{itemize}

In the CTMC model, the backoff time and the transmission duration are computed from an exponential distribution. Moreover, due to the continuous-time property of the CTMCs, packet collisions due to the same random backoff computation are not captured. For further details on the usage and construction of CTMCs for characterizing IEEE 802.11 WLANs, we refer the interested reader to the work in~\cite{barrachina2019dynamic}.

\begin{figure}[ht!]
    \centering    
    \includegraphics[width=\columnwidth]{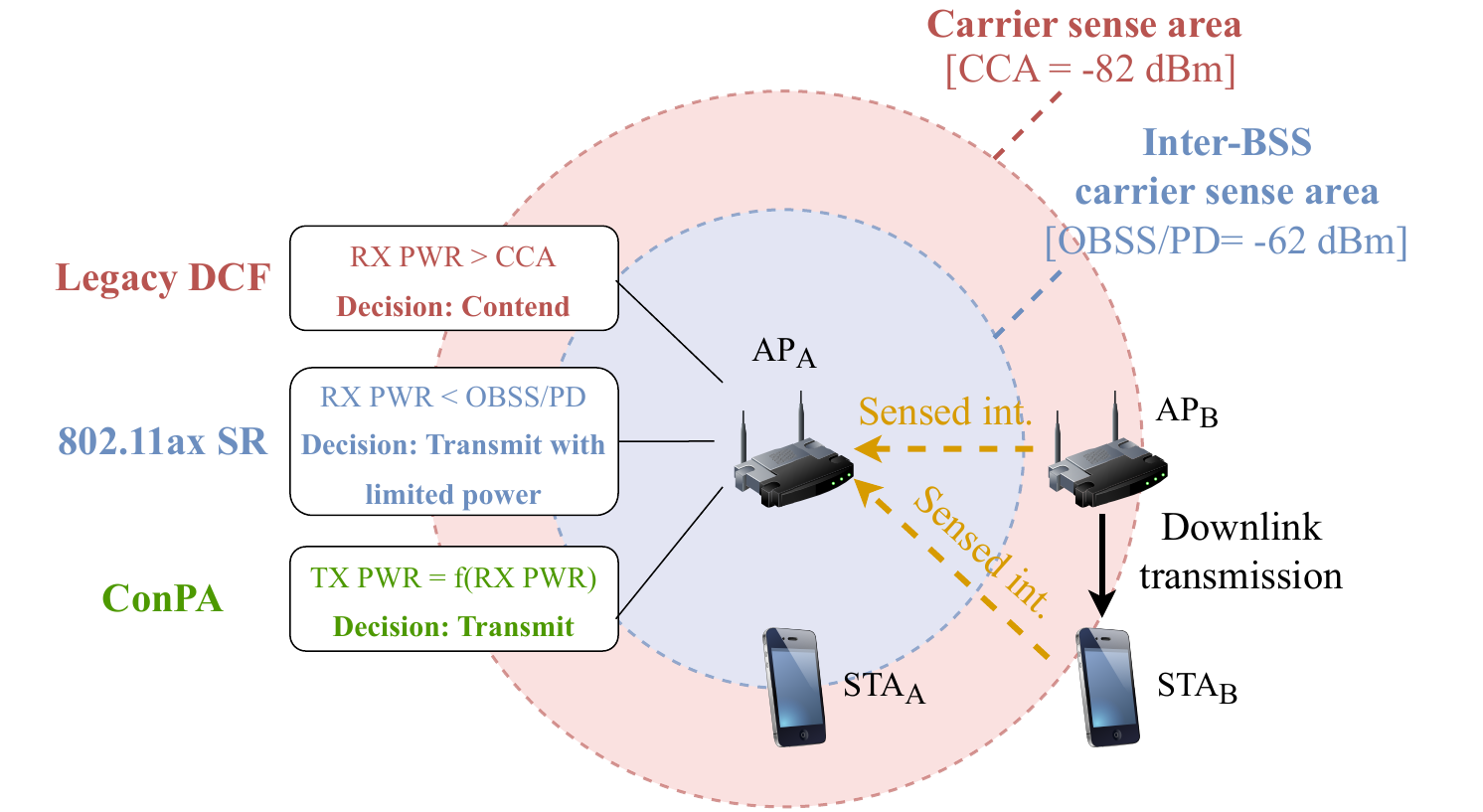}
    \caption{Channel access policies in an exemplary deployment.}
    \label{fig:example_scenario_ctmn}
\end{figure}

To showcase a CTMC representation of a particular deployment implementing DCF, 802.11ax SR, and ConPA, we focus on the symmetric scenario in Fig.~\ref{fig:example_scenario_ctmn}, where two BSSs potentially interfere with each other. The red and blue areas in Fig.~\ref{fig:example_scenario_ctmn} illustrate the carrier sense range and inter-BSS carrier sense area of AP$_A$ when applying the DCF and 802.11ax SR, respectively. Any detected transmission from a device falling in these areas would trigger channel contention according to DCF or 802.11ax SR (depending on the case). In Fig.~\ref{fig:example_scenario_ctmn}, if using DCF, AP$_A$ would contend for the channel when AP$_B$ is transmitting (and vice versa). With 802.11ax SR, in contrast, AP$_A$ would transmit upon limiting its transmit power based on the OBSS/PD threshold applied to disregard AP$_B$'s transmissions, since AP$_B$ is out of AP$_A$'s inter-BSS carrier sense area. Finally, the concept of contention no longer holds for ConPA, but transmissions are subject to a transmit power limitation based on sensed interference.

\begin{figure}[ht!]     
    \begin{center}
    \begin{tikzpicture}[->, >=stealth', auto, semithick, node distance=3cm]
    \tikzstyle{every state}=[fill=white,draw=black,thick,text=black,scale=1]
    \node[state]    (0)    {$\emptyset$};
    \node[state]    (A)[above right of=0, xshift=-0.5cm, yshift=-0.5cm]   {$A$};
    \node[state]    (B)[below right of=0,  xshift=-0.5cm, yshift=0.5cm]   {$B$};
    \node[state] (AB)[right of=A, xshift=-0.5cm, yshift=1cm, fill=ForestGreen!30] {$AB^\diamondsuit$};
    \node[state] (BA)[right of=B, xshift=-0.5cm, yshift=-1cm, fill=ForestGreen!30] {$A^\diamondsuit B$};    
    \node[state] (ABSR)[below of=AB, xshift=0.5cm, yshift=1.25cm, fill=NavyBlue!30] {$AB^\spadesuit$};
    \node[state] (BASR)[below of=ABSR, yshift=1.25cm, fill=NavyBlue!30] {$A^\spadesuit B$};
    \node[state] (ABDCF)[right of=ABSR, xshift=-1.5cm, yshift=-0.9cm, fill=red!20] {$AB$};
    \path
    (0) edge[bend left] node{\scriptsize$\alpha\lambda^A$} (A)
        edge[bend right] node{\scriptsize$\alpha\lambda^B$}  (B)
    (A) edge[bend right] node{\scriptsize$\mu^A$} (0)
    (B) edge[bend left] node{\scriptsize$\mu^B$} (0)
    (A) edge[ForestGreen, left, above] node{\scriptsize$\alpha\lambda^B$} (AB)
    (B) edge[ForestGreen, right, above] node{\scriptsize$\alpha\lambda^A$} (BA)   
    (AB) edge[ForestGreen, right, below] node{\scriptsize$\mu^{B^\diamondsuit}$} (A)
    (BA) edge[ForestGreen, left, below] node{\scriptsize$\mu^{A^\diamondsuit}$} (B)
    (AB) edge[right, color=ForestGreen, at start,anchor=north west] node{\scriptsize$\mu^A$} (B)
    (BA) edge[left, color=ForestGreen, at start,anchor=south west] node{\scriptsize$\mu^B$} (A)
    (A) edge[NavyBlue, dashed, left, below, pos=0.3] node{\scriptsize$\alpha\lambda^B$} (ABSR)
    (B) edge[NavyBlue, dashed, right, above, pos=0.3] node{\scriptsize$\alpha\lambda^A$} (BASR)  
    (ABSR) edge[NavyBlue, dashed, right, pos=0.1, below] node{\scriptsize$\mu^A$} (0)
    (BASR) edge[NavyBlue, dashed, left, above, pos=0.1] node{\scriptsize$\mu^B$} (0)
    ;
    \end{tikzpicture}
    \end{center}        
    \caption{CTMC representation of the 2-BSS deployment for the different considered channel access mechanisms. States and transitions for DCF (unaccessible in the showcased deployment), 802.11ax states, and ConPA states, are illustrated in \hlc[red!20]{red}, \hlc[NavyBlue!30]{blue}, and \hlc[ForestGreen!30]{green}, respectively.}
    \label{fig:ctmc}
\end{figure}
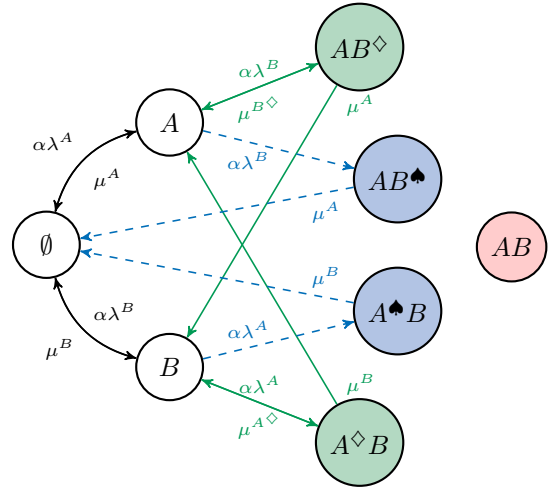

Figure~\ref{fig:ctmc} shows the CTMC characterizing the deployment in Fig.~\ref{fig:example_scenario_ctmn} for each channel access mechanism. Notice that the common states for all the mechanisms are illustrated in white, while specific states and transitions of each mechanism are represented in \hlc[red!20]{red} (DCF), \hlc[NavyBlue!30]{blue} (802.11ax SR), and \hlc[ForestGreen!30]{green} (ConPA). In DCF, BSSs A and B alternate the access to the channel due to the CCA policy, so state $AB$ is unreachable. In 802.11ax SR, both BSSs can transmit simultaneously, provided that the BSS leveraging the OBSS/PD-based SR TXOP applies a transmit power limitation (indicated by the superscript $\spadesuit$). Notice that backward transitions from states in which simultaneous transmissions occur (e.g., $AB^\spadesuit$) are ruled by the TXOP of the BSS winning the channel access first (e.g., $A$ in state $AB^\spadesuit$). Finally, in ConPA, BSSs can access the channel at any time, provided that they apply $f$. The transmit power limitation applied by a given BSS is indicated by the superscript $\diamondsuit$. To avoid recursive states in characterizing ConPA, we assume that the transmit power used by BSSs is independent in each state and subject to the hierarchy established when accessing the channel in an orderly manner. For instance, when BSS A accesses the channel first, there is a transition to state A. A subsequent transmission byBSS B is subject to a transmit power limitation based on the power sensed from BSS A, thus leading to state $AB^\diamondsuit$. But as soon as BSS A finishes its transmission, the chain transitions to state B, where BSS B does not need to apply a transmit power limitation anymore because the interference from BSS A has ceased.

\section{System Model}
\label{sec:csr}

\subsection{Channel Model}
\label{sec:channel_model}

Both the transmit and the sensed power of a given transmission are assumed to be fixed in each state in the CTMC. The power received by a receiver device, $\mathcal{P}_{rx}$, is affected by the following log-distance path loss model with obstacles~\cite{wilhelmi2021spatial}:
\begin{equation}
    \text{PL}(d_{n,m}) = \text{PL}_{0} + 10\nu \log_{10}(d_{n,m}) + \frac{\sigma}{2} + \frac{\omega}{2} \frac{d_{n,m}}{10},
    \label{eq:pathloss}
\end{equation}
where $\text{PL}_{0}$ is the path loss intercept, $\nu$ is the path loss exponent, $d_{n,m}$ is the distance between the transmitter $m$ and the receiver $n$, $\sigma$ is the shadowing factor, and $\omega$ is the objects' attenuation factor.

\subsection{Packet Transmission Model}
\label{sec:transmission_model}

Having Request-to-Send/Clear-to-Send (RTS/CTS) in place, the duration of a data transmission, upon being either successful ($T_{\mathrm{succ}}$) or failed ($T_{\mathrm{fail}}$), is given by
\begin{equation}
\begin{cases}
   T_{\mathrm{succ}} = & T_\text{RTS} + 3\cdot T_\text{SIFS} + T_\text{CTS} + T_\text{DATA} + \\&T_\text{ACK} + T_\text{DIFS} + T_{\mathrm{e}}, \\
    T_{\mathrm{fail}} = & T_\text{RTS} + T_\text{SIFS} + T_\text{CTS} + T_\text{DIFS} + T_{\mathrm{e}},
\end{cases}
\label{eq:transmission_duration}
\end{equation}
where control frames (RTS, CTS, and ACK) and standard intervals (SIFS, DIFS) have a fixed duration (see~\cite[Table~B.6]{wilhelmi2021spatial}). As for the duration of the data segment ($T_\text{DATA}$), it depends on the total number of data frames aggregated in an A-MPDU during the TXOP duration limit, $\text{TXOP}_{\max}$, and the Modulation and Coding Scheme (MCS) used. MCS selection is done on a per-CTMC-state basis and is based on the RSSI perceived at the STA. Accordingly, the maximum MCS leading to a maximum Packet Error Rate (PER) of 10\% is selected. We consider the MCS options included in the IEEE 802.11ax, where modulations range from BPSK with a coding rate of 1/2 to 1024-QAM with a coding rate of 5/6.

We consider that packet losses are caused by weak SINR, which results from \textit{i)} insufficient transmit power, \textit{ii)} high path loss attenuation, and/or \textit{iii)} excessive neighboring interference. We assume that a packet can be decoded at a receiver $n$ if its SINR, $\gamma^{(n)}$, is above a Capture Effect (CE) threshold, $\gamma_{\textrm{CE}}$. Notice that, since CTMC states are independent, packet errors are accounted only in the states where $\gamma^{(n)} < \gamma_{\textrm{CE}}$.

\subsection{Traffic Model}

Only downlink transmissions are considered, which is required to keep the complexity of the CTMC model low. Nevertheless, as shown in Section~\ref{sec:transmission_model}, uplink control frames are considered to compute the total duration of a data transmission. The characterization of the traffic load is done through the buffer status, which is regulated by a parameter $\alpha\in [0,1]$ that indicates the probability of having packets ready when the backoff timer expires. This parameter is used to regulate the forward transitions rates in the CTMC.
Traffic is generated following a Poisson process, where the time between generated packets is exponentially distributed with mean $1/l$, where $l$ is the packet generation rate.

\subsection{Key Performance Indicators}
\label{sec:kpi}

To assess the performance of ConPA and compare it with the baseline methods, we define the following metrics:
\begin{itemize}
    \item \textbf{Throughput, $\Gamma^{(n)}$ (bps):} The throughput of BSS $n$ is defined as the total data bits transmitted with respect to the total time. It is computed as $\Gamma^{(n)} = \sum_{s \in \mathcal{S}} \pi_s \mathbbm{1}_{\{\gamma^{(n)}_s \geq \gamma_{\textrm{CE}}\}}  \big( N^{(n)}_{a,s} \, L_D \, R_s^{(n)} \big)$, where $N^{(n)}_{a,s}$ is the number of aggregated data frames by BSS $n$ in state $s$, $L_D$ is the length of an individual data payload frame, and $R_s^{(n)}$ is the data rate used.
    \item \textbf{Airtime, $\rho^{(n)}$ [\%], and Airtime efficiency $\hat{\rho}^{(n)}$ [\%]:} The airtime is the total percentage of the time a BSS accesses the channel, computed as $\rho^{(n)} = 100 \times \sum_{s \in \mathcal{S}} \pi_s \mathbbm{1}_{\{n\in \Phi_s\}}$, where $\Phi_s$ indicates the set of BSSs active in state $s$. The airtime efficiency focuses on the percentage of airtime with successful transmissions.
    \item \textbf{Mean MCS, $\overline{\Omega}^{(n)}$ [unitless]:} The mean MCS index used in transmissions, computed as $\overline{\Omega}^{(n)} = (\rho^{(n)}/100)^{-1} \sum_{s \in \mathcal{S}} \Omega_{s}^{(n)}$.
    \item \textbf{Mean SINR, $\overline{\gamma}^{(n)}$ [dB]:} The mean SINR experienced in transmissions, computed as $\overline{\gamma}^{(n)} = (\rho^{(n)}/100)^{-1} \sum_{s \in \mathcal{S}} \gamma_{s}^{(n)}$.
\end{itemize}

\section{Performance Evaluation}
\label{sec:performance_evaluation}

\begin{table}[t!]
\centering
\caption{Evaluation parameters.}
\label{tbl:simulation_parameters}
\resizebox{.8\columnwidth}{!}{%
\begin{tabular}{@{}clc@{}}
\toprule
Parameter & \multicolumn{1}{c}{\textbf{Description}} & \textbf{Value} \\ \midrule
$K$ & Num. of evaluated random drops & 1000 \\
$B$ & Transmission bandwidth & 80 MHz \\
$F_c$ & Carrier frequency & 6 GHz \\
$SUSS$ & Single-user spatial streams & 2 \\
$\mathcal{P}_{tx,\max}$ & Max. transmit power & 20 dBm \\
$\mathcal{P}_{\min}$ & Min. transmit power used in ConPA & 1 dBm \\
$\mathcal{P}^\text{Noise}$ & Noise power & -95 dBm \\
$G^\text{TX/RX}$ & Transmitter/receiver antenna gain & 0/0 dBi \\
$PL_0$ & Loss at the reference dist. & 5 dB \\
$\nu$ & Path-loss exponent & 4.4 \\
$\sigma$ & Shadowing factor & 9.5 dB\\
$\omega$ & Obstacles factor & 30 dB\\
$\gamma_{\textrm{CE}}$ & Capture effect threshold & 10 dB \\
$\text{CW}$ & Contention window & 16 \\
$L_{D}$ & Length of data packets & 12000 bits \\ 
$\text{TXOP}_{\max}$ & TXOP duration limit & 5,484 ms\\ 
\bottomrule
\end{tabular}%
}
\end{table}

To evaluate ConPA, we focus on different scenarios of $N\in\{2,4\}$ BSSs using the same frequency channel with one AP and one STA each. Each AP is placed at the center of a squared cubicle with side length $D\in\{2,4,8\}$~m and its associated STA is randomly deployed within the same cubicle's area.\footnote{The different cubicles are located next to each other, forming grids of $1\times2$ and $2\times2$, respectively.} The evaluation parameters are summarized in Table~\ref{tbl:simulation_parameters}.

\begin{figure}[ht!]
    \centering     
    \includegraphics[width=\linewidth]{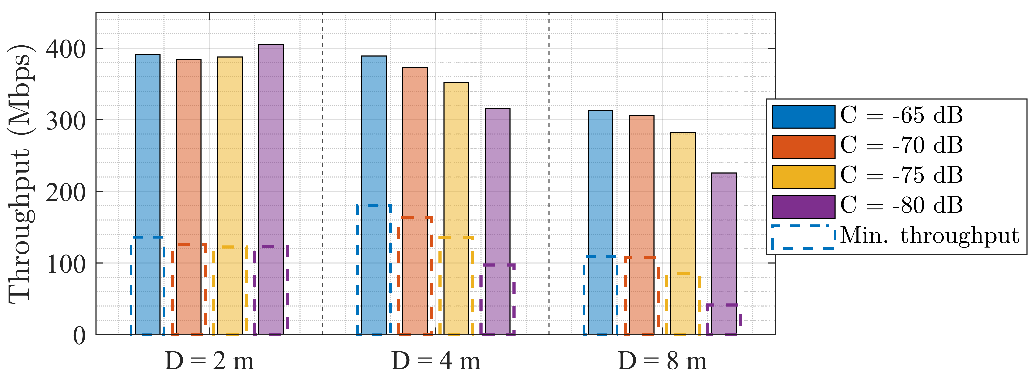}
     \caption{Average throughput achieved by $N=4$ APs (with $\alpha=1$) by ConPA for $D\in \{2,4,8\}$ m, $\text{C}\in \{-65, -70, -75, -80\}$, and $\alpha=1$.} 
    \label{fig:fig1}
\end{figure}

First, we evaluate the effect of parameter $\text{C}$ on the performance of ConPA in Fig.~\ref{fig:fig1}, which shows the average throughput across all the evaluated deployments with $N=4$ BSSs, for $\text{C} = \{-65, -70, -75, -80\}$ dB and $D\in\{2,4,8\}$~m (representing small, medium, and large scenarios). The results in Fig.~\ref{fig:fig1} include the average of both the mean throughput (solid bars) and the minimum throughput (dashed bars) obtained among the BSSs in each random deployment. As shown, $\text{C}$ leads to different performance depending on the size of the scenario. Low $\text{C}$ values ($\text{C}=-80$~dB) provide higher throughput in small scenarios ($D=2$~m), but they perform poorly as $D$ increases because the imposed transmit power reduction is excessive. In contrast, higher values of $\text{C}$ (e.g., $\text{C}=-65$~dB) lead to higher throughput at medium and large scenarios ($D=4$~m and $D=8$~m). Nevertheless, their throughput degradation at short distances is minor. For that reason, we use $\text{C}=-65$~dB for the remainder of the paper.

\begin{figure*}[t!]
    \centering
    \begin{subfigure}{\linewidth}
        \centering
        \includegraphics[width=.9\linewidth]{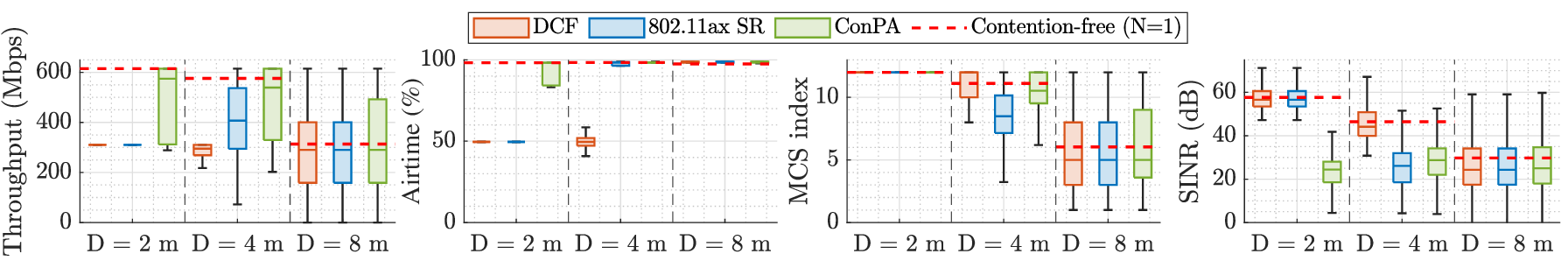}
        \caption{$N = 2$ BSSs.}
        \label{fig:2a}
     \end{subfigure}
     \hfill
     \begin{subfigure}{\linewidth}
        \centering
        \includegraphics[width=.9\linewidth]{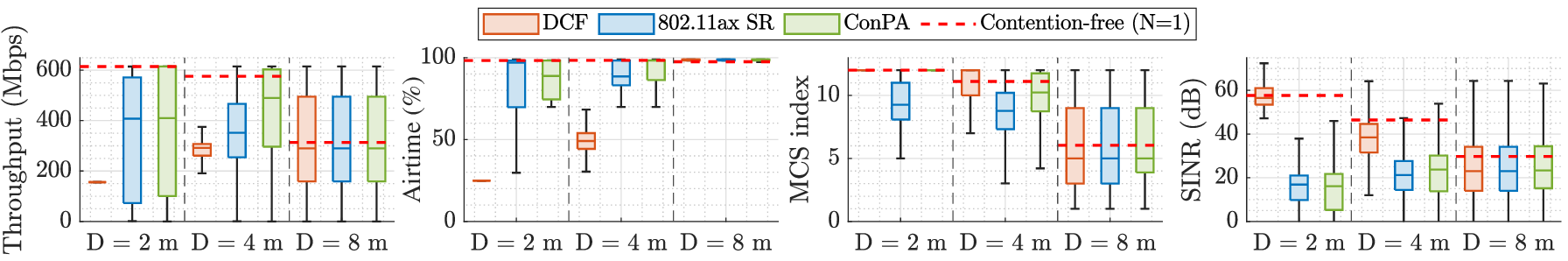}
        \caption{$N = 4$ BSSs.}
        \label{fig:2b}
     \end{subfigure}
     \caption{Boxplot of various performance indicators achieved by each mechanism for $N \in \{2, 4\}$ BSSs, $D\in \{2,4,8\}$ m, and $\alpha=1$. The baseline performance of contention-free DCF (i.e., $N=1$ BSS) is shown by the red dashed lines.}
    \label{fig:fig2}
\end{figure*}

Next, we compare the performance of ConPA against the DCF and 802.11ax SR baselines. For that, Fig.~\ref{fig:fig2} provides the boxplots\footnote{A boxplot shows the median (line inside the box), the worst/best 25\% performance quartiles (top and bottom edges of the box, respectively), and the maximum/minimum values (whisker lines outside the box).} of different metrics---throughput, airtime, average MCS, and average SINR---at different scenarios with $N=2$ BSSs (Fig.~\ref{fig:2a}, on the top) and $N=4$ BSSs (Fig.~\ref{fig:2b}, on the bottom). We evaluate the performance based on the deployment size, for $D\in \{2,4,8\}$ m. 

Starting with the left-most plots in Fig.~\ref{fig:fig2}, it is proved that ConPA provides the highest throughput among all the mechanisms and for all the considered cubicle sizes, with up to a 76\% improvement on the median of the throughput for $N=2$~BSSs and $D=2$~m. At greater values of $D$, the benefits of ConPA are reduced as the physical distance between devices increases, meaning that devices no longer need to compete for the channel. The gain in throughput achieved by ConPA can be explained through the airtime utilization, as shown in the second plot from the left in Fig.~\ref{fig:fig2}. Devices implementing ConPA perform transmissions bypassing channel contention, resulting in up to 98\% more airtime. Finally, MCS and SINR figures on the right in Fig.~\ref{fig:fig2} show how the baselines DCF and 802.11ax SR allow performing transmissions with higher quality (e.g., DCF obtains up to 71\% more SINR than ConPA for $N=4$~BSSs and $D=2$~m). Despite the quality of the transmissions using ConPA is lower, it still achieves better throughput than the baselines, and this is thanks to the great spatial reuse properties exhibited by the mechanism. This can be explained by considering that the baselines are conceived to avoid collisions, so that any transmissions are held with potentially less interference. For $D=4$~m, however, 802.11ax SR experiences a performance anomaly, as a result of sensing the interference at the transmitter and not the receiver~\cite{wilhelmi2021spatial}.

To delve more into the quality of the simultaneous transmissions enabled by ConPA, Table~\ref{tab:airtime_efficiency} provides the average airtime efficiency for each studied scenario. With a higher correlation with the previous results, we observe that ConPA leads to lower airtime efficiency, especially at low distances (up to a 52\% decrease for $N=4$~BSSs and $D=2$~m). Nevertheless, the gains in the total airtime shown in Fig.~\ref{fig:fig2} allow ConPA to provide higher throughput when compared to the baselines.

\begin{table}[t!]
\centering
\caption{Airtime efficiency ($\hat{\rho}$) achieved by each mechanism for $N \in \{2, 4\}$ BSSs, $D\in \{2,4,8\}$ m, and $\alpha=1$.}
\label{tab:airtime_efficiency}
\resizebox{\columnwidth}{!}{%
\begin{tabular}{@{}cccc|ccc@{}}
\toprule
\multirow{2}{*}{} & \multicolumn{3}{c}{$N = 2$ BSSs} & \multicolumn{3}{c}{$N = 4$ BSSs} \\ \cmidrule(l){2-7} 
 & $D = 2$~m & $D = 4$~m & $D = 8$~m & $D = 2$~m & $D = 4$~m & $D = 8$~m \\ \cmidrule(r){1-7}
DCF & 100\% & 100\% & 82.09\% & 100\% & 77.54\% & 74.65\% \\
802.11ax SR & 100\% & 87.41\% & 82.09\% & 52.24\% & 66.59\% & 74.65\% \\
ConPA & 74.61\% & 88.26\% & 86.91\% & 48.27\% & 74.96\% & 81.54\% \\ \bottomrule
\end{tabular}%
}
\end{table}

\begin{figure*}[t!] 
    \centering
    \includegraphics[width=.7\textwidth]{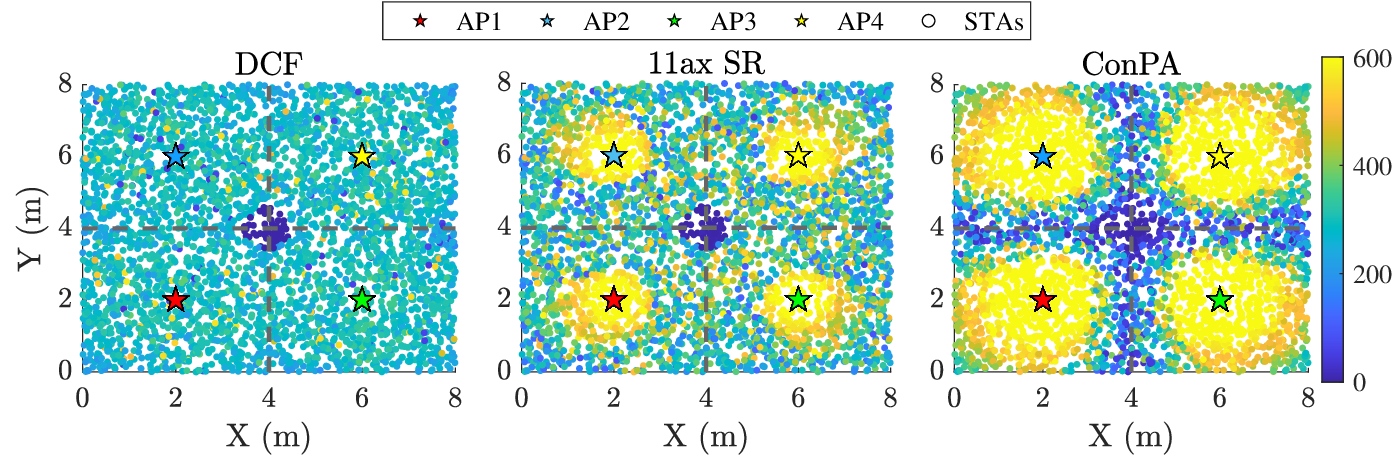}
    \caption{Throughput (in Mbps) obtained in each random deployment, for $N=4$~BSSs, $D = 4$~m, and $\alpha = 1$.}
    \label{fig:throughput_map_4aps_4m}
\end{figure*}

To provide a more intuitive understanding on ConPA's behavior when compared to DCF and 802.11ax, Fig.~\ref{fig:throughput_map_4aps_4m} visually shows the throughput achieved in each random deployment evaluated for $D=4$~m and $N=4$ BSSs. The plots show the physical location of the STAs within their cubicles plus the throughput obtained in each random drop using color intensities. As shown, ConPA provides great throughput improvements for STAs near their APs while leading to slightly worse performance than the baselines for some STAs located at the cell edges. This underscores the role of ConPA, which sacrifices SINR in favor of airtime, which is precisely what LBT neglects. Note that such a role might be particularly useful in use cases like sub-networks with very stringent latency requirements~\cite{berardinelli2021extreme}.

\begin{figure}[t!]
    \centering
    \includegraphics[width=.9\linewidth]{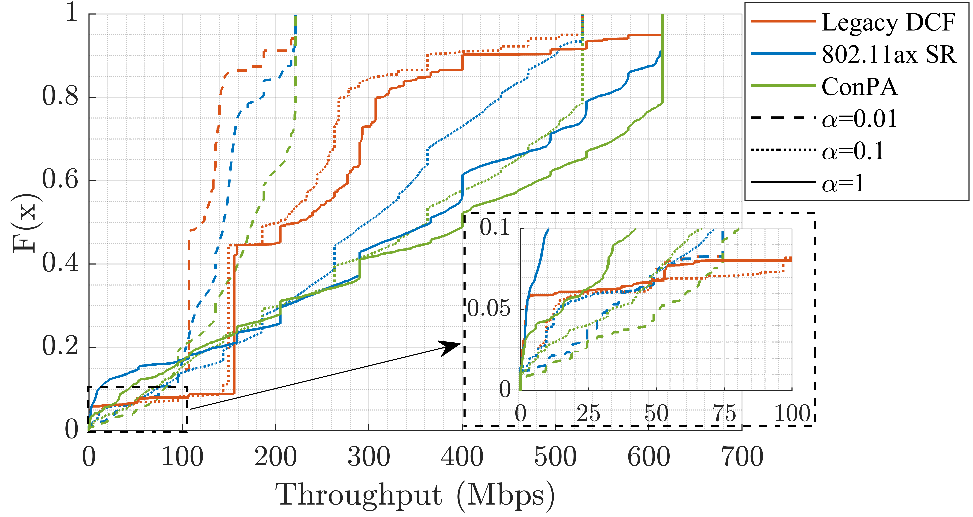}
    \caption{Empirical CDF of the throughput obtained by each mechanism for each load value $\alpha \in \{0.01, 0.1, 1\}$, for $N=4$ BSSs and $D\in\{2, 4, 8\}$~m.}
    \label{fig:load_cdf}
\end{figure}

To conclude, in Fig.~\ref{fig:load_cdf}, we look into the gains of ConPA under different traffic load regimes, represented through $\alpha=\{0.01, 0.1, 1\}$ to depict from low load to saturated traffic conditions. As shown, regardless of the load, ConPA succeeds in providing more throughput than the rest of the methods in almost all cases, being the highest gains under saturated traffic. When it comes to worst-case scenarios from 0-100 Mbps (depicted by the zoomed portion of the plot in Fig.~\ref{fig:load_cdf}), we observe that DCF outperforms ConPA in certain cases for $\alpha=\{0.1, 1\}$, indicating the need to alternate transmissions in some situations (potentially involving STAs located at the edge of the cubicles) when the load increases.

\section{Conclusions}
\label{sec:conclusions}

In this paper, we proposed and evaluated ConPA, a novel channel access mechanism whereby wireless devices can bypass channel contention to better support the reliability and low latency requirements of next-generation applications. To showcase the gains of ConPA, we provided an analytical model based on CTMCs and evaluated ConPA against legacy DCF and 802.11ax SR. Our results evidence the ability of ConPA to reuse the space, thus allowing simultaneous transmissions with reduced impact on their quality, even with a high network density. Unlike overprotective methods based on LBT, which prioritize interference-free transmissions over latency, ConPA is designed to improve airtime. Our results show that ConPA improves the throughput of the baselines by up to 76\%.

\bibliographystyle{IEEEtran}
\bibliography{bib}

\end{document}